\documentclass[journal]{vgtc}                

\ifpdf
  \pdfoutput=1\relax                   
  \usepackage{graphicx}                
  \DeclareGraphicsExtensions{.pdf,.png,.jpg,.jpeg} 
\else
  \usepackage{graphicx}                
  \DeclareGraphicsExtensions{.eps}     
\fi%

\graphicspath{{figures/}{pictures/}{images/}{./}} 

\usepackage{microtype}                 
\PassOptionsToPackage{warn}{textcomp}  
\usepackage{textcomp}                  
\usepackage{mathptmx}                  
\usepackage{times}                     
\usepackage{cite}                      
\usepackage{tabu}                      
\usepackage{booktabs}                  

\onlineid{0}

\vgtccategory{Research}
\vgtcpapertype{please specify}

\title{Projector Pixel Redirection Using Phase-Only Spatial Light Modulator}

\author{Haruka Terai, Daisuke Iwai, and Kosuke Sato}
\authorfooter{
\item
 Haruka Terai, Daisuke Iwai, and Kosuke Sato are with Graduate School of Engineering Science, Osaka University
\item
 Daisuke Iwai is with PRESTO, Japan Science and Technology Agency.
}

\shortauthortitle{}

\abstract{In projection mapping from a projector to a non-planar surface, the pixel density on the surface becomes uneven. This causes the critical problem of local spatial resolution degradation. We confirmed that the pixel density uniformity on the surface was improved by redirecting projected rays using a phase-only spatial light modulator.%
} 


\renewcommand{\manuscriptnotetxt}{}

\nocopyrightspace

\vgtcinsertpkg

\begin{document}


\firstsection{Introduction}

\maketitle


Projection mapping is a technology for superimposing an image onto a three-dimensional object with an uneven surface using a projector. It changes the appearance of objects such as color and texture, and is used in various fields, such as medical~\cite{00000658-201806000-00024} or design support~\cite{8797923}. Projectors are generally designed so that the projected pixel density is uniform when an image is projected on a flat screen. When an image is projected on a non-planar surface, the pixel density becomes uneven, which leads to significant degradation of spatial resolution locally.

Previous works applied multiple projectors to compensate for the local resolution degradation. Specifically, projectors are placed so that they illuminate a target object from different directions. For each point on the surface, the previous techniques selected one of the projectors that can project the finest image on the point. In run-time, each point is illuminated by the selected projector~\cite{1634329,10.1145/2159516.2159518,Nagase2011}. Although these techniques successfully improved the local spatial degradation, their system configurations are costly and require precise alignment of projected images at sub-pixel accuracy. To the best of our knowledge, few attempts have been done thus far to solve the problem by a single projector approach.

This study aims to overcome the technical limitation by making the pixel density uniform on a projection surface to be used for dynamic projection mapping, we apply a phase-only spatial light modulator (PSLM) as a dynamic free-form lens to redirect projected rays. Because adaptive ray redirection is possible owing to the real-time control capability of PSLM, this technique potentially works even for dynamic projection mapping scenarios in which a projector and/or a surface are moved.

\section{Proposed Method}

This paper proposes a pixel redirection method using PSLM that can be used for dynamic projection mapping. First, the pixel locations that are spatially uniform on the surface is computed. Next, a phase image for PSLM is computed, which is necessary to achieve the desired pixel locations. Then, the generated phase image is used for pixel redirection by PSLM to achieve a uniform pixel density on the surface.

This section describes the configuration of the proposed projection system and the phase image generation method.

\subsection{Phase-Only Spatial Light Modulator (PSLM)}

\begin{figure}[t]
  \centering
  \includegraphics[width=0.98\hsize]{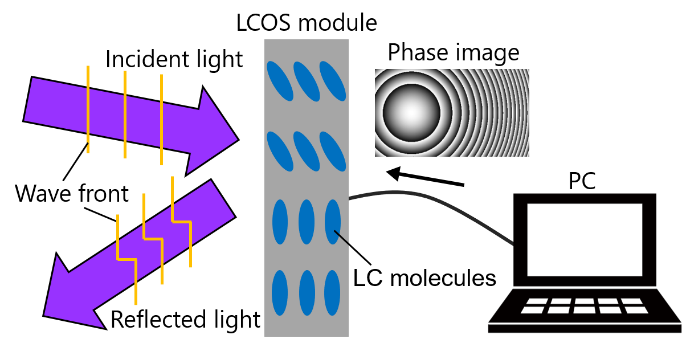}
  \caption{Phase modulation principle by PSLM-LCOS.}
	\label{fig:1}
\end{figure}

A PSLM is a device that spatially modulates the phase of an incident light wave. We use a reflective PSLM consisting of a liquid crystal on silicon (LCOS) module. When a two-dimensional control data, called a phase image, is provided to the PSLM, it changes the direction of LC molecules according to the control electricity, as shown in~\autoref{fig:1}. The refractive index changes as the LC molecules rotate. This changes the optical path length of the light passing through the LC, resulting in an optical phase delay. When the input light is linearly polarized so that the polarization direction is the same as the alignment direction of the LC molecules, only the phase is modulated. Thus, we can control the wavefront of the input light by controlling the spatial distribution of the phase of light by providing a tailored phase image to the PSLM.

\subsection{Proposed System}

\begin{figure}[t]
  \centering
  \includegraphics[width=0.6\hsize]{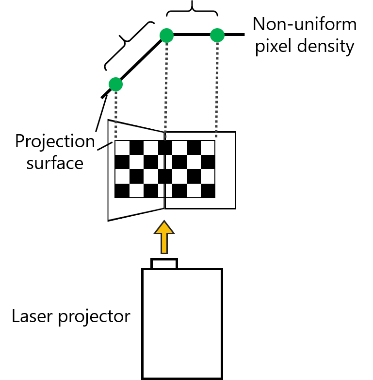}\\
  \makebox[0.98\hsize][c]{\raisebox{0.5ex}[0ex][0ex]{\footnotesize (a) Normal projection}}\\\ \\
  \includegraphics[width=0.98\hsize]{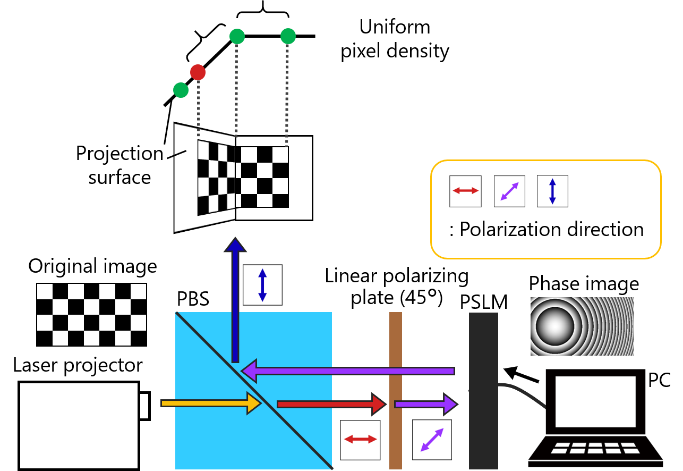}\\
  \makebox[0.98\hsize][c]{\raisebox{0.5ex}[0ex][0ex]{\footnotesize (b) Pixel redirection projection}}
  \caption{Proposed method.}
	\label{fig:2}
\end{figure}

\autoref{fig:2} shows a schematic diagram of the proposed system. It consists of a polarization beam splitter (PBS), a linear polarizing plate, a laser scanning projector, and a PSLM. The PBS is used to prevent the projection of unmodulated light. The linearly polarizing plate is an optical element that transmits only linearly polarized light whose axis corresponds to the transmission axis direction. It is applied to make the polarization direction of the input light to the LCOS module the same as the alignment direction of the LC molecules and to enable phase-only modulation. Because each pixel of the laser projector can hit on the LCOS module without overlaps between adjacent pixels, the laser projector is employed as the light source to enable the independent modulation of each projector pixel.

\subsection{Phase Image Computation}

\begin{figure}[t]
  \centering
  \includegraphics[width=0.98\hsize]{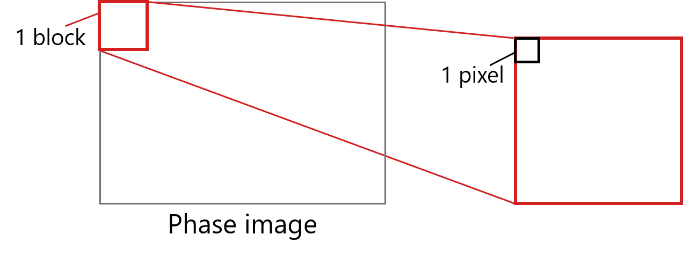}
  \caption{Phase image segmentation.}
	\label{fig:3}
\end{figure}

To generate the phase image that achieves the desired projected pixel distribution, we first conducted an experiment to measure the amount of pixel shift while changing the phase image consisting of spatially repeated pattern. Next, based on this measurement, the relationship between the phase image pattern and the amount of pixel shift is stored in a lookup table. Then, as shown in \autoref{fig:3}, the phase image is divided into blocks corresponding to a single pixel of the projector. The derived relationship is used to obtain the brightness variation between adjacent pixels of the phase image pattern to achieve the desired pixel arrangement in each block. Finally, based on this value, the phase image is generated by determining the pixel values of the phase image to have C1 continuity throughout the phase image.

\section{Experiment}

\begin{figure}[t]
  \centering
  \includegraphics[width=0.98\hsize]{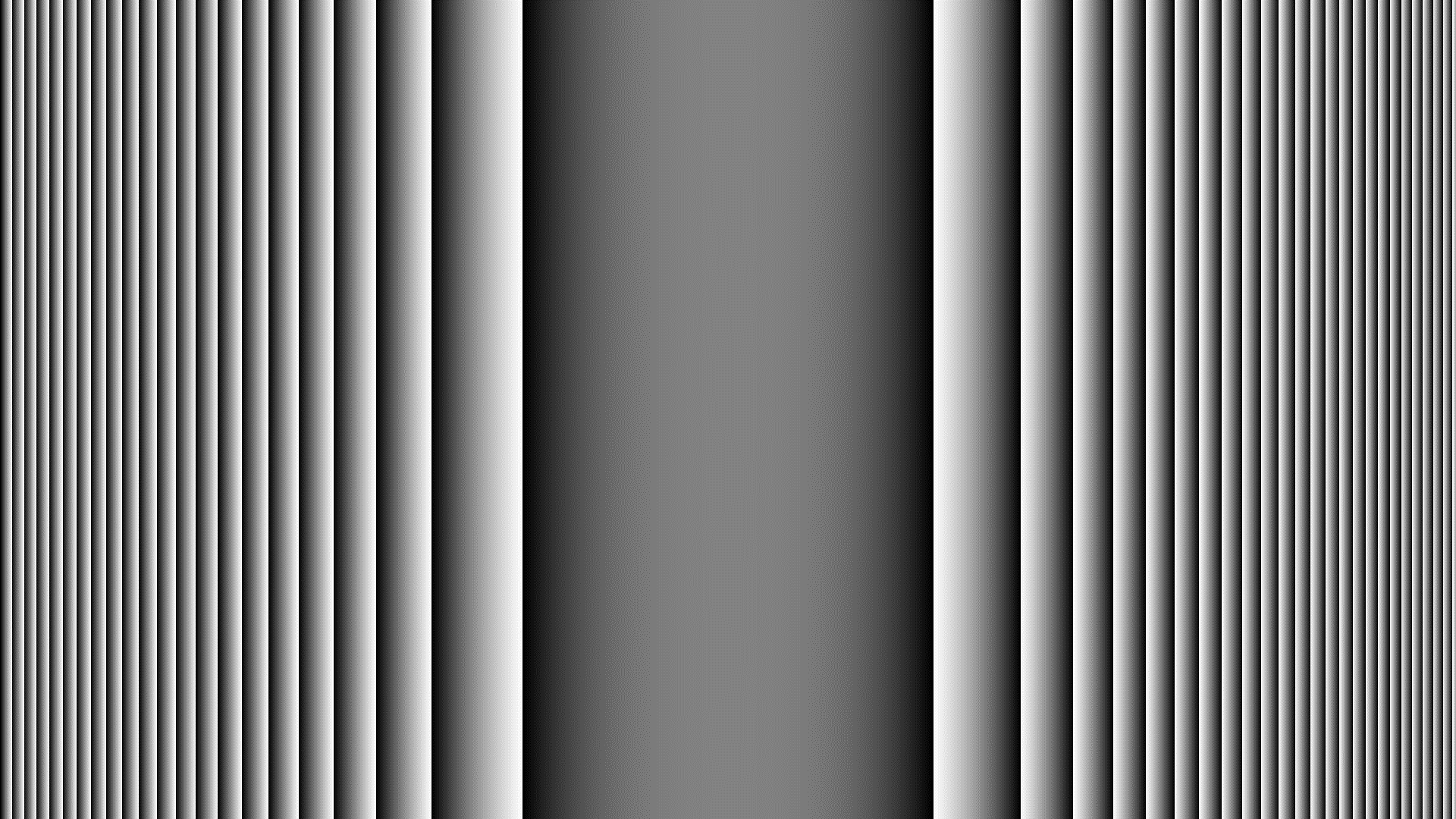}\\
  \makebox[0.98\hsize][c]{\raisebox{0.5ex}[0ex][0ex]{\footnotesize (a) Phase image}}\\\ \\
  \includegraphics[width=0.98\hsize]{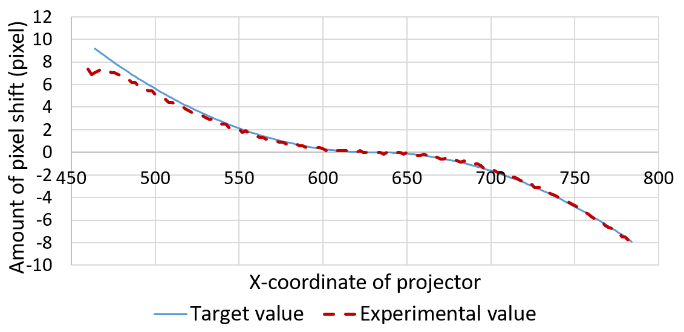}\\
  \makebox[0.98\hsize][c]{\raisebox{0.5ex}[0ex][0ex]{\footnotesize (b) Relationship between X-coordinate and pixel shift amount}}
  \caption{Experimental result.}
	\label{fig:4}
\end{figure}

We conducted an experiment to verify whether the proposed phase image generation method and phase modulation system can achieve the target pixel shift. We set a target pixel shift in the horizontal direction. Then our method generated the phase image to realize this shift. \autoref{fig:4}(a) shows the obtained phase image. \autoref{fig:4}(b) is a graph showing the target pixel shift and the actual pixel shift in the experiment, with the right direction as positive. This result shows that we could achieve the pixel shift close to the target value.

\begin{figure*}[t]
  \centering
  \includegraphics[width=0.48\hsize]{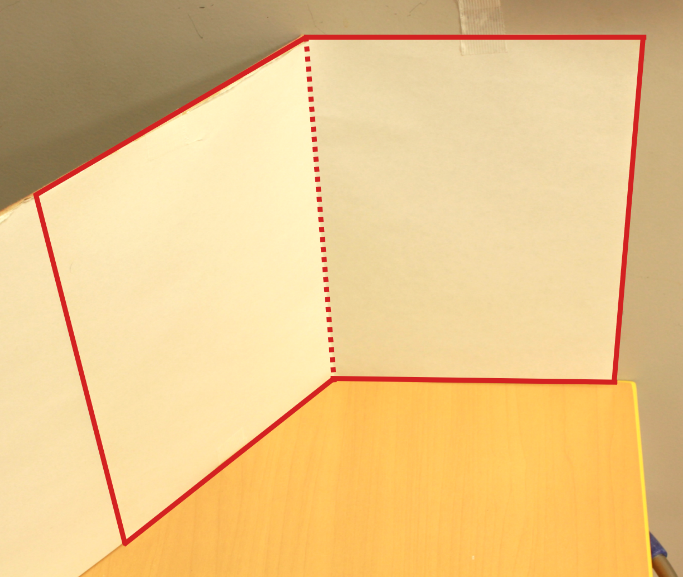}
  \includegraphics[width=0.48\hsize]{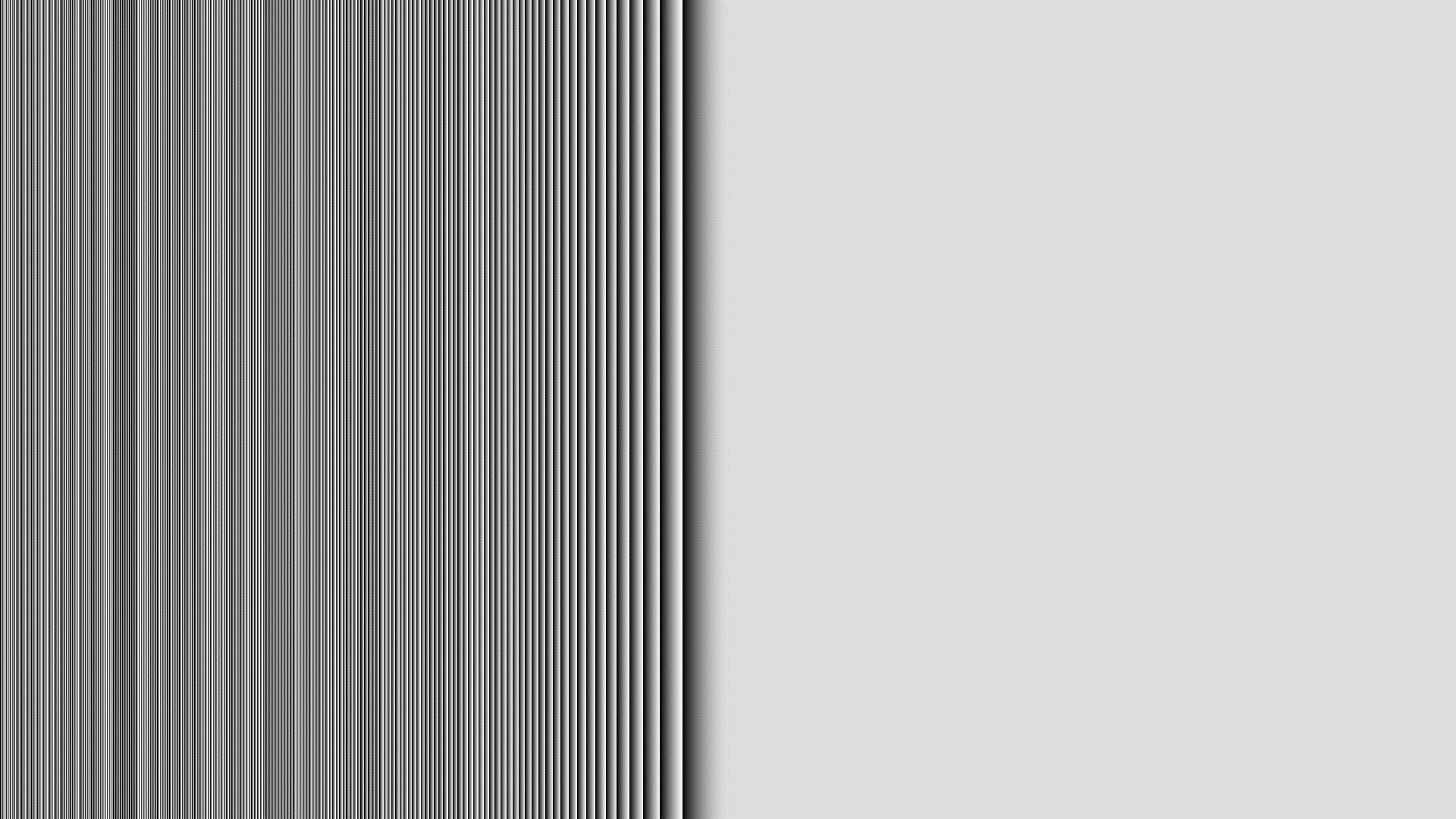}\\
  \makebox[0.48\hsize][c]{\raisebox{0.5ex}[0ex][0ex]{\footnotesize (a) Projection plane (surrounded in red line)}}
  \makebox[0.48\hsize][c]{\raisebox{0.5ex}[0ex][0ex]{\footnotesize (b) Phase image}}\\\ \\
  \includegraphics[width=0.48\hsize]{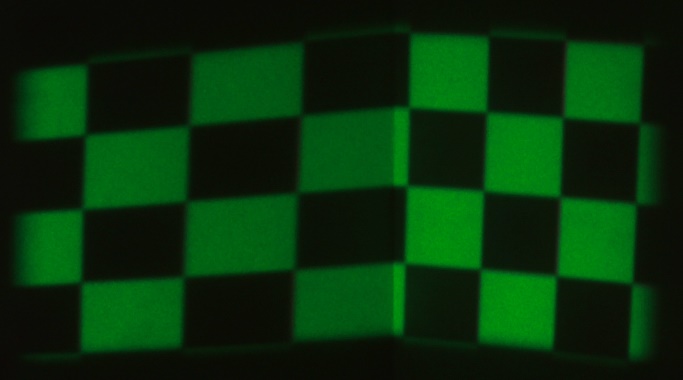}
  \includegraphics[width=0.48\hsize]{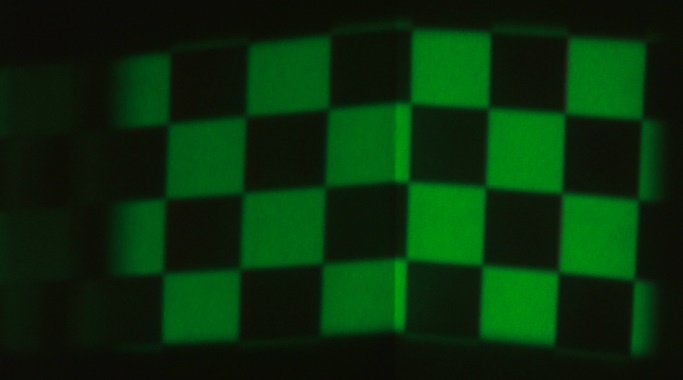}\\
  \makebox[0.48\hsize][c]{\raisebox{0.5ex}[0ex][0ex]{\footnotesize (c) Projection result (before modulation)}}
  \makebox[0.48\hsize][c]{\raisebox{0.5ex}[0ex][0ex]{\footnotesize (d) Projection result (after modulation)}}
  \caption{Pixel density uniformization.}
	\label{fig:5}
\end{figure*}

\autoref{fig:5} shows a pixel density uniformization result on a projection surface bent at a center. Before the uniformization, the checker pattern on the projection surface on the left side of the scene stretched to the left. To the contrary, when our pixel redirection method applied, the pixel density uniformity was improved.

\section{Conclusion}

In this paper, we conducted a study aiming at the pixel density uniformization on a non-planar projection surface by redirecting projector pixels using a phase-only spatial light modulator (PSLM). We confirmed that the pixel density uniformity on the projection surface was improved using the proposed method.

\acknowledgments{
This work was supported by JST, PRESTO Grant Number JPMJPR19J2.}

\bibliographystyle{abbrv-doi}

\bibliography{template}
\end{document}